\pdfoutput=1
\documentclass[prl,superscriptaddress,showpacs,twocolumn]{revtex4}

\usepackage{graphicx}

\usepackage{graphicx,amsfonts}

\usepackage{epsfig,amsmath}

\usepackage{verbatim}

\begin{document}

\newcommand{\atanh}
{\operatorname{atanh}}

\newcommand{\ArcTan}
{\operatorname{ArcTan}}

\newcommand{\ArcCoth}
{\operatorname{ArcCoth}}

\newcommand{\Erf}
{\operatorname{Erf}}

\newcommand{\Erfi}
{\operatorname{Erfi}}

\newcommand{\Ei}
{\operatorname{Ei}}

\newcommand{\sgn}{{\mathrm{sgn}}}
\newcommand{\rme}{{\mathrm{e}}}
\newcommand{\rmd}{{\mathrm{d}}}
\def\be{\begin{equation}}
\def\ee{\end{equation}}

\def\bea{\begin{eqnarray}}
\def\eea{\end{eqnarray}}

\def\e{\epsilon}
\def\l{\lambda}
\def\d{\delta}
\def\o{\omega}
\def\cb{\bar{c}}
\def\Li{{\rm Li}}

\title{Free-energy distribution of the directed polymer at high temperature}

\author{Pasquale Calabrese}
\affiliation{Dipartimento di Fisica dell'Universit\`a di Pisa and INFN, 56127 Pisa Italy}
\author{Pierre Le Doussal}
\email{ledou@lpt.ens.fr}
\affiliation{CNRS-Laboratoire de Physique
Th{\'e}orique de l'Ecole Normale Sup{\'e}rieure, 24 rue Lhomond, 75231
Paris Cedex, France}
\author{Alberto Rosso}
\affiliation{CNRS-Universit\'e Paris-Sud, LPTMS, UMR8626-B\^at 100,91405 Orsay Cedex, France}

\date{\today}

\begin{abstract}

We study the directed polymer of length $t$ in a random potential with fixed endpoints in dimension $1+1$ in the continuum 
and on the square lattice, by analytical and numerical methods. 
The universal regime of high temperature $T$ is described, upon scaling `time' $t \sim T^5/\kappa$ and 
space $x = T^3/\kappa$ (with $\kappa=T$ for the discrete model) by a continuum model with $\delta$-function disorder correlation.
Using the Bethe Ansatz solution for the attractive boson problem, we obtain all positive integer moments of the partition function. The lowest cumulants of the free energy are predicted at small time and found in agreement with numerics. 
We then obtain the exact expression at any time for the generating function of the free energy distribution, in terms of a Fredholm determinant. At large time we find that it crosses over to the Tracy Widom distribution (TW) which describes the fixed $T$ infinite $t$ limit. 
The exact free energy distribution is obtained for any time 
and compared with very recent results on growth and exclusion models. 
\end{abstract}

\maketitle

The directed polymer (DP) in a random potential provides the simplest example of a
glass phase induced by quenched disorder \cite{directedpoly} and has numerous applications, e.g.
vortex lines \cite{vortex}, domain walls \cite{lemerle}, biophysics \cite{hwa}.
It is closely related to much studied growth models in the KPZ class \cite{KPZ}, such as 
asymmetric exclusion processes (ASEP) \cite{spohnTASEP,MallickWASEP}, and to Burgers turbulence \cite{Burgers}. 
It belongs to the broader class of disordered elastic manifolds, known to exhibit statistically
scale invariant ground states. Within the functional RG (FRG) \cite{frg} these were described
in a dimensional expansion by $T=0$ fixed points, where the ratio temperature/disorder 
is irrelevant and scales with internal size with exponent $-\theta$.

Exact results were obtained in dimension $d=1+1$  \cite{directedpoly}. Johansson 
proved \cite{Johanssonzeta,Johansson2000} that (i) the minimal energy path of length $t$ on a square lattice with fixed endpoints has transverse roughness
$x \sim t^{\zeta}$ with $\zeta=\frac{2}{3}$ (ii) the fluctuation of the ground state energy grows as
$t^\theta$ with $\theta=\frac{1}{3}$ and its scaled distribution coincides with the one of the smallest 
eigenvalue of a hermitian random matrix, the GUE Tracy Widom (TW)
distribution \cite{TW1994}. The TW distribution was found in many
other related models, polynuclear growth \cite{spohn2000}, TASEP \cite{spohnTASEP},
random subsequences \cite{Baik1999,Majumdar2005} and others \cite{Majumdar2004,TW2001,loc}.
The unifying concept of determinantal space-time process and edge scaling 
was studied to account for such universality \cite{spohnreview}. An exact result for
the space-time scaling function of the two-point correlator of the height in KPZ 
was obtained \cite{spohnKPZ0}. 

On the other hand, in $d=1+1$ the model can be mapped onto the quantum mechanics of 
$n$ attractive bosons in the limit $n=0$, where $t$ plays the role of (imaginary) time. 
It can be solved with the Bethe Ansatz (BA) for
$\delta$-function interactions. Until now only the ground state 
energy $E_0(n)$ was studied,  i.e. the limit $t \to \infty$ first. 
Pioneering attempts at its direct analytical continuation at $n=0$ for a system of 
transverse size $L=\infty$ led to scaling behavior \cite{kardareplica,bo}, but not to free energy
distribution. The possible dominance of rare events is also a problem of the infinite system. 
From continuation at fixed $L$, Brunet and Derrida obtained \cite{BD} the large 
deviation function for the fluctuations of the free energy $\delta F \sim L^{1/2}$ of the DP 
on the cylinder. This, however, is different from the distribution of free energy at fixed $t$, 
which requires a summation over excited states. Also, it has been a longstanding question whether
the $\delta$-function model captures the low-$T$ physics. For example, the 
FRG suggests otherwise, i.e. that some structure of the disorder correlator matters.
In fact, Brunet's BA result \cite{BrunetDiffusion} for the diffusion coefficient around the cylinder, $D \sim (\kappa T)^{-1/2}$, 
does not reproduce the expected finite $T=0$ limit. This remains to be reconciled with the standard argument
of a single glass phase controlled by the $T=0$ fixed point, which suggests a single
universality. These issues are also outstanding for Burgers and KPZ growth,
and related ASEP models also solvable via Bethe Ansatz \cite{spohnTASEP,MallickWASEP}.

In this Letter we perform the sum over excited states and obtain the exact expression for the free energy distribution
from the Bethe Ansatz. The lowest cumulants of the free energy are computed at small time 
and checked with numerics. The generating function of the free energy distribution is obtained at any time as a Fredholm
determinant. At large time it shows that the free energy distribution crosses over to the Tracy Widom distribution.
The probability distribution of the free energy is also obtained at any time. Our study, started independently, 
parallels a recent work by Dotsenko and Klumov \cite{dotsenko}. Although we agree on the starting sum over 
states in \cite{dotsenko}, our analysis allows to recover TW.  
Finally, we discuss the behaviour of the amplitudes as a function of the temperature. 

Let us recall the various definitions of the DP model.
(i) {\it continuum model:} the partition sum with fixed endpoints
is defined by the path integral $Z=Z(0,0,t)$ with:
\be \label{zdef} 
Z(x,y,t) = \int_{x(0)=x}^{x(t)=y}  Dx e^{- \frac{1}{T} \int_0^t d\tau [ \frac{\kappa}{2}  (\frac{d x}{d\tau})^2  + V(x(\tau),\tau) ]} 
\ee
for a given realization of the centered gaussian random potential $V(x,t)$ of correlator
$\overline{V(x,t) V(x',t)} = \delta(t-t') R(x-x')$. Upon replication of (\ref{zdef}), disorder averaging and Feynman-Kac formula, one finds that the disorder averages ${\cal Z}_n := \overline{Z(x_1,y_1,t)..Z(x_n,y_n,t)}$ satisfy $\partial_t {\cal Z}_n = - H^{rep}_n {\cal Z}_n$ with Hamiltonian:
\bea \label{general}
&& H^{rep}_n =  - \frac{T}{2 \kappa} \sum_{i=1}^n \partial_{x_i}^2 
- \frac{1}{2 T^2} \sum_{ij} R(x_i-x_j) 
\eea
and attractive interaction $- R(x)/T^2$. It is known from FRG that to describe
low-$T$ physics one must retain some features of $R(x)$, i.e. that it is a decaying function 
on the correlation scale $r_f$. At high $T$, however, if one defines 
\begin{eqnarray} \label{resc}
x=T^3 \kappa^{-1} \tilde x \quad , \quad t=2 T^5 \kappa^{-1} \tilde t
\end{eqnarray}
in coordinates $\tilde x$ and $\tilde t$ one has $Z= \int D\tilde x e^{-S}$, with $S= \int d\tilde t [ \frac{1}{4} (\partial_{\tilde t} \tilde x)^2  + W(\tilde x, \tilde t) ]$
where $\overline{ W(\tilde x , \tilde t) W(\tilde x' , \tilde t') }= \tilde R(\tilde x - \tilde x') \delta(\tilde t - \tilde t')$
and $\tilde R(\tilde x) = 2 T^3 \kappa^{-1} R(T^3 \kappa^{-1} \tilde x)$. When
$T \to \infty$ one has $\tilde R(z) \to 2 \bar c \delta(z)$ with $\bar c=\int du R(u)$. Hence 
 in that limit the general model (\ref{general}), expressed in the coordinates $\tilde x$, $\tilde t$ becomes the Lieb-Liniger (LL) model \cite{ll},
 i.e.  $t H^{rep}_n = \tilde t H_{LL}|_{x_i \to \tilde x_i}$ of Hamiltonian:
\be
H_{LL} = -\sum_{j=1}^n \frac{\partial^2}{\partial {x_j^2}} 
+ 2 c \sum_{1 \leq  i<j \leq n} \delta(x_i - x_j),
\label{LL}
\ee
where $c=-\bar c$ is the interaction parameter. The LL model is thus the {\it simultaneous limit} $T,x,t \to \infty$ of the DP problem with $\tilde x$ and $\tilde t$ fixed. It should also describe the region where $T^3 (\bar c \kappa)^{-1} \gg r_f$, i.e. $T \gg T_{dep}$, the crossover ``thermal depinning" temperature, well known in vortex physics \cite{vortex,footnote2,mueller2001}. For $T>T_{dep}$ the thermal fluctuations average out partially the disorder and $R$ can be replaced by a $\delta$ correlator, while for $T<T_{dep}$ the finite range of $R(x)$ is essential for the physics of pinning. One outstanding question is whether, for a fixed $T \gg T_{dep}$ the LL model describes the system all the way as it flows to the $T=0$ fixed point, or whether fixed but large $\tilde t$ is a distinct limit from $T=0$. 

(ii) {\it discrete model:}

For the numerics we define the partition sum $\tilde Z_{i,j}=\sum_{\gamma} e^{- \beta \sum_{(r,s) \in \gamma} V_{r,s}}$
over all paths $\gamma$ directed along the diagonal on a square lattice, with only $(1,0)$ or $(0,1)$ moves, starting in $(0,0)$ and 
ending in $(i,j)$, where the $\tilde V_{r,s}$ are i.i.d. random site variables. Introducing ``time" $\hat t=i+j$ and space
$\hat x = \frac{i-j}{2}$, $Z_{\hat x,\hat t}=\tilde Z_{i,j}$ satisfies:
\be \label{discrete}
Z_{\hat x,\hat t+1}=(Z_{\hat x-\frac{1}{2},\hat t}+Z_{\hat x+ \frac{1}{2},\hat t}) e^{-\beta V_{\hat x,\hat t+1}}
\ee
with 
$Z_{\hat x,0}=\delta_{\hat x,0}$. We are interested in the free energy $F=-T \ln Z$ with $Z=Z_{\hat x=0,\hat t}$
of paths of length $\hat t$ returning to the origin. At $T=0$ and for a geometric distribution
Johansson proved \cite{Johansson2000} that the ground state energy $F_{T=0} \approx e_0 \hat t + \sigma \omega \hat t^{1/3}$ with $Prob(\omega>-s)=F_2(s)$ the TW distribution \cite{TW1994}. 

In the high $T$ limit this model maps onto the continuum one (\ref{zdef}) with $\kappa=4 T$ and 
$\delta$-function correlation when expressed in the variables (\ref{resc}), i.e. $\tilde x=4 \hat x/T^2$ and
$\tilde t=2 \hat t/T^4$. Following \cite{BrunetThesis} one checks that $Z(\tilde x,\tilde t)/\bar Z$ with 
$\bar Z=2^{\hat t} e^{\frac{1}{2} \beta^2 \hat t}$ is given by the LL model (\ref{LL}) and we find that
unit gaussian on the lattice corresponds to $\bar c=1$.

We now use the Bethe Ansatz solution of the LL model (\ref{LL}). The 
moments of (\ref{zdef}), expressed in $\tilde x$, $\tilde t$ coordinates
(we drop the tilde below except when stated otherwise) can be expressed as a quantum
mechanical expectation:
\bea
&& \overline{Z^n} =\langle x_0 \dots x_0 |e^{-t H_{LL}}| x_0 \dots x_0 \rangle\,, \\
&& = \sum_\mu \frac{|\langle x_0 \dots x_0|\mu\rangle|^2}{||\mu ||^2} e^{-t E_\mu}\,.
\label{mom}
\eea
i.e. all $n$ replica start and end at $x_0=0$,
and we used the resolution of the identity in terms of the eigenstates $|\mu\rangle$ of $H_{LL}$ 
of energies $E_\mu$. The crucial observation, which makes the calculation tractable, is that 
only symmetric (i.e. bosonic) eigenstates contribute to this average. 
The eigenfunctions are superpositions of plane waves \cite{ll}
$\Psi_\mu= F[\lambda] \sum_P A_P \prod_{j=1}^n e^{i \l_{P_\ell} x_\ell}$ 
over all permutations $P$ of the rapidities $\l_j$ and $F[\lambda]=\prod_{n \geq \ell > \ell' \geq 1} \frac{\lambda_{\ell}-\lambda_{\ell'}}{\sqrt{(\lambda_{\ell}-\lambda_{\ell'})^2 + c^2}}$. The coefficients 
$A_P=\prod_{n \geq \ell > k \geq 1} (1- \frac{i c ~\text{sgn}(x_\ell - x_k))}{\lambda_{P_\ell} - \lambda_{P_k}})$
are functions of the two-particle scattering 
phase shifts obtained from (\ref{LL}), and periodicity of the wavefunction requires
the set of rapidities $\{ \l \}$ to be solution to the Bethe equations. Major simplifications
occur in this complicated equations in the attractive case $\cb > 0$ for $L=\infty$. 
They have complex, i.e. bound states solutions \cite{m-65}.
A general eigenstate is built by partitioning the $n$ particles into a set of $n_s$ 
bound-states formed by $m_j \geq 1$ particles with $n=\sum_{j=1}^{n_s} m_j$. 
The rapidities associated to a bound-state form a regular pattern in the complex plane which is called 
{\it string} $\l^{j, a}=k_j +\frac{i\cb}2(j+1-2a)+i\d^{j,a}$. 
Here, $a = 1,...,m_j$ labels the rapidities within the string.  
$\d^{j,a}$ are deviations which fall off exponentially with system size $L$.
Perfect strings (i.e. with $\d=0$) are exact eigenstates in the limit $L\to\infty$ for arbitrary $n$.
Such eigenstates have definite momentum $K_\mu=\sum_{j=1}^{n_s} m_j k_j$ and energy $E_\mu=\sum_{j=1}^{n_s} (m_j k_j^2-\frac{\cb^2}{12} m_j(m_j^2-1))$. The ground-state corresponds to a single $n$-string with $k_1=0$. The string-states are commonly believed to be a complete set, although a rigorous proof is still missing.

To evaluate (\ref{mom}), we first obtain the string wavefunction at coinciding point 
$\langle 0 \cdots 0 |\mu\rangle =  \Psi_\mu(0,..0) = n! F[\lambda]$.
The computation of the norms $||\mu||$ of string states is more involved, but was solved
in the context of algebraic Bethe Ansatz \cite{cc-07} (see also \cite{kk}). It reads:
\begin{eqnarray}
&& ||\mu||^2 = \frac{n!  (L \bar{c})^{n_s} }{(\bar{c})^{n}} \frac{F[\lambda]^2}{
 \Phi[k,m]}  \prod_{j=1}^{n_s} m_j^2  \\
&& \Phi[k,m]=
\prod_{\substack{1\leq i<j\leq n_s}} 
\frac{(k_i-k_j)^2 +(m_i-m_j)^2 c^2/4}{(k_i-k_j)^2 +(m_i+m_j)^2 c^2/4} \nonumber 
\end{eqnarray}
Expressing the sum over states in (\ref{mom}) as all partitioning of 
$n$ particles into $n_s$ strings and using that for $L\to\infty$ the string
momenta $m_j k_j$ correspond to free particles \cite{cc-07}, i.e. 
$\sum_{k_j} \to m_j L \int \frac{dk_j}{2 \pi}$ we obtain \cite{us}:
\begin{eqnarray}  \label{partsum}
&& 
\overline{\hat Z^n} = \sum_{n_s=1}^n \frac{n!}{n_s! (2\pi \bar c)^{n_s}} \sum_{(m_1,\dots m_{n_s})_n} 
\\&& 
\int \prod_{j=1}^{n_s} \frac{d k_j}{m_j}
 \Phi[k,m]
\prod_{j=1}^{n_s}e^{m_j^3  \frac{\cb^2 t}{12}- m_j k_j^2 t}\,, \nonumber 
\end{eqnarray} 
where $(m_1,\dots m_{n_s})_n$ stands for all the partitioning of $n$ such that $\sum_{j=1}^{n_s} m_j=n$
with $m_j \geq 1$. We defined $Z=\bar{c} e^{- \frac{c^2 t}{12}} \hat Z$, a trivial shift in the free energy (we drop
the hat below). This equation agrees with the one in \cite{dotsenko}, although our derivation was made simpler 
by using results from algebraic Bethe ansatz. 

\begin {figure}
\begin{center}
\includegraphics[width=8cm]{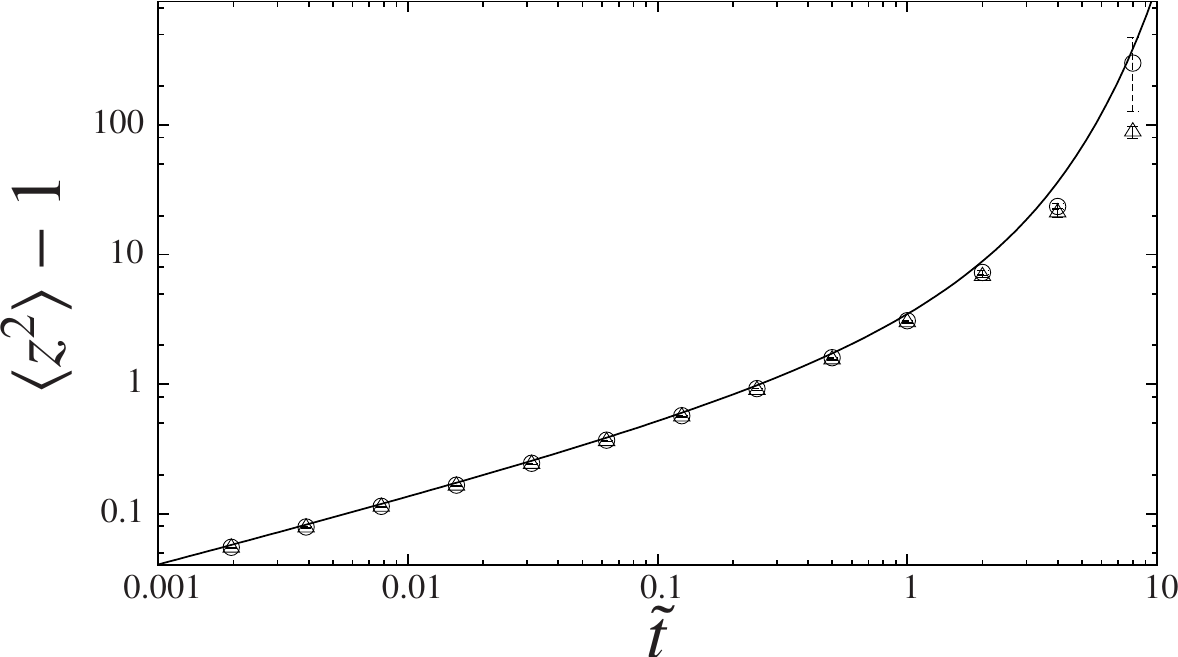}
\end{center}
\caption{$\overline{z^2}-1$ ($4 ~10^6$ samples) for $\hat t=128$ (triangle), $\hat t= 256$ (circle) function of $\tilde t$ compared to formula (\ref{z2ana}) with $\bar c=1$. } 
\label{figz2}
\end {figure}

\begin {figure}
\begin{center}
\includegraphics[width=8cm]{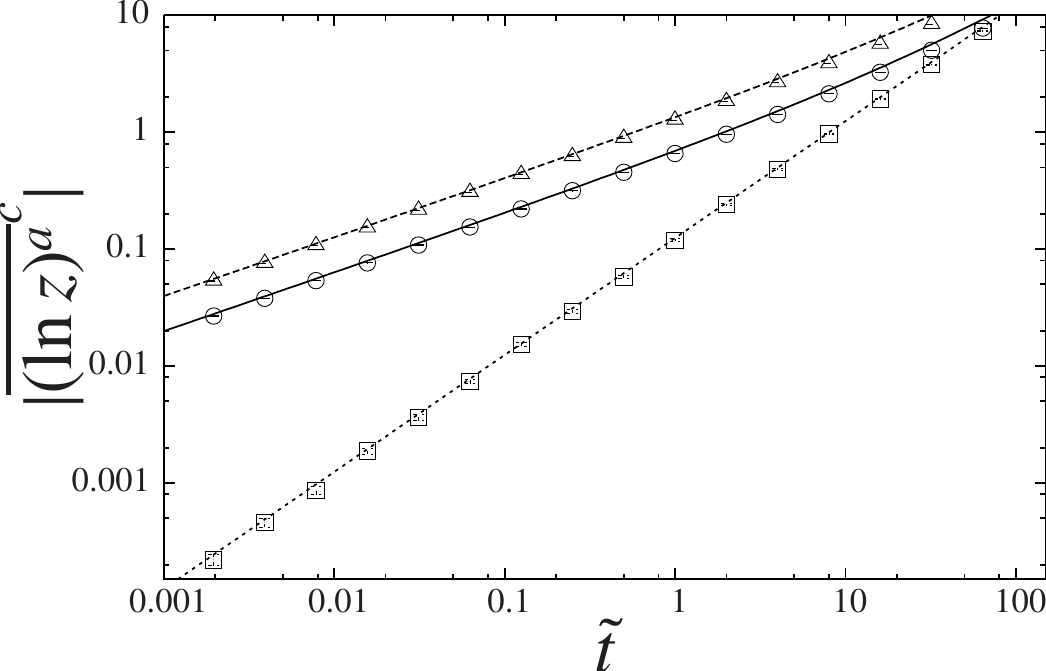}
\end{center}
\caption{From top to bottom the cumulants ($4 ~10^6$ samples) $\overline{(\ln z)^2}^c$ (dashed line, triangle),
$- \overline{(\ln z)}$ (solid line, circle), and $\overline{(\ln z)^3}^c$ 
(dotted line, square) for $\hat t=256$ as compared with the 
the analytical formula (\ref{logcum}) with $\bar c=1$.}
\label{figlogcum}
\end {figure}

This formula first leads to prediction at small time. As in \cite{dotsenko} we define the dimensionless parameter:
\begin{eqnarray}  
&& \lambda = (c^2 \tilde t/4)^{1/3} 
\end{eqnarray} 
and $z=Z/\overline{ Z}$. Tedious calculation then yields:
\begin{eqnarray} \label{z2ana}
&& \overline{z^2} =  1 + \sqrt{2 \pi} \lambda^{3/2} e^{2 \lambda^3} (1 +  {\rm erf}(\sqrt{2} \lambda^{3/2}))  \\
&& \overline{\ln z} = - \sqrt{\frac{\pi}{2}} \lambda^{3/2} + (\frac{32 \pi}{9 \sqrt{3}} -2 - \frac{3 \pi}{2}) \lambda^3 + .. \nonumber \\
&& \overline{(\ln z)^2}^c =  \sqrt{2 \pi} \lambda^{3/2} + (4 + 5 \pi - \frac{32 \pi}{3 \sqrt{3}}) \lambda^3 + .. \nonumber \\
&& \overline{(\ln z)^3}^c =  (\frac{32}{3 \sqrt{3}} - 6) \pi \lambda^3 + ..   \label{logcum}
\end{eqnarray}
The skweness of the distribution of $\ln z$ is thus $\gamma_1^{\ln z}  \approx   \frac{16 \sqrt{3} - 27}{9} (2 \pi)^{1/4} \lambda^{3/4} \sim t^{1/4}$ at small time. The skewness for the free energy $F= - T \ln Z$ is thus $\gamma_1^F = - \gamma_1^{\ln z}$ and
{\it negative}.  Fig. \ref{figz2} and Fig. \ref{figlogcum} show that the agreement with numerics is excellent {\it with no free parameter}. This is a non trivial test that the LL model is valid here and that the starting formula (\ref{partsum}) is correct. 

\begin {figure}
\begin{center}
\includegraphics[width=8cm]{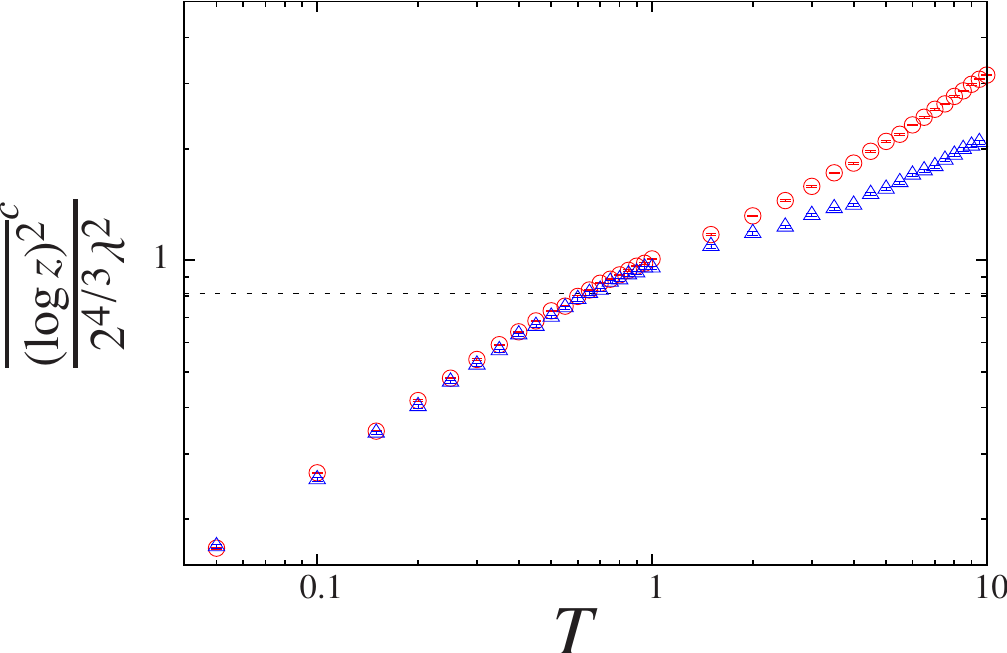}
\end{center}
\caption{$\overline{(\ln z)^2}^c/(2^{4/3} \lambda^2)$ plotted as a function of $T$, for increasing polymer length $\hat t$. Triangles correspond to $\hat t=4096$, Circles to $\hat t=256$ and the dotted line to the TW variance $0.81319..$. Averages are performed over 20000 samples.}
\label{figf2}
\end {figure}

To study any $\lambda$, we avoid the explicit $n=0$ limit
by introducing the generating function of the distribution $P(f)$ of the 
scaled free energy $F=T \lambda f$:
\begin{eqnarray}  
&& g(x) = 1 + \sum_{n=1}^\infty \frac{(- e^{\lambda x})^n}{n!} \overline{Z^n} = 
\overline{ \exp( - e^{\lambda (x - f)} ) } 
\end{eqnarray} 
from which $P(f)$ is immediately extracted at $\lambda \to \infty$:
\begin{eqnarray}  
&& \lim_{\lambda \to \infty} g(x) = \overline{ \theta(f-x) } = Prob(f > x) 
\end{eqnarray} 
At finite $\lambda$ the probability distribution of the free energy 
can also be extracted from $g(x)$ using a Borel transform, equivalently
described as follows. One writes (formally) $Z=Z_0 \tilde Z$ where $Z_0=e^{\lambda u_0}$ is a 
positive random variable independent from $\tilde Z$, with an exponential distribution 
$P_0(Z_0)=e^{-Z_0}$ (i.e. $\lambda u_0$ has a unit Gumbel distribution) such that $\overline{Z^n}=n! \overline{\tilde Z^n}$. The
distribution $P(\tilde Z)$ of the variable $\tilde Z=e^{\lambda u}$ is obtained from the cut in the grand canonical partition function ${\cal Z}(z)=\sum_{n=0}^\infty z^{-n} \overline{\tilde Z^n}= \overline{\frac{z}{z-\tilde Z}}$ as \cite{footnotepos}:
\begin{equation} \label{extract}
 z P(z)= \frac{1}{\pi}  {\rm Im} {\cal Z}(z+i \epsilon) = \frac{1}{\pi} {\rm Im}
g(e^{\lambda x} \to - \frac{1}{z-i \epsilon}) 
\end{equation} 
with $\epsilon=0^+$. 

The constraint $\sum_{i=1}^{n_s} m_i=n$ in (\ref{partsum}) can then be relaxed,
and rescaling $k_j \to k_j/t^{1/2}$, it leads to:
\begin{eqnarray} \label{defg}
g(x) = 1 +  \sum_{n_s=1}^\infty \frac1{n_s!}  Z(n_s,x) 
\end{eqnarray}
as an expansion in the number of strings with \cite{footnoteconv}:
\begin{widetext}
\begin{eqnarray}  
&& Z(n_s,x) =  \sum_{m_1,\dots m_{n_s}=1}^\infty \frac{(-1)^{\sum_j m_j} }{(4 \pi \lambda^{3/2})^{n_s}}
 \prod_{j=1}^{n_s} \int \frac{d k_j}{m_j}  
 \prod_{\substack{1\leq i<j\leq n_s}} 
\frac{(k_i-k_j)^2 + (m_i-m_j)^2 \lambda^3}{(k_i-k_j)^2 + (m_i+m_j)^2 \lambda^3} \prod_{j=1}^{n_s}
e^{\frac{1}{3} \lambda^3 m_j^3 -  m_j k_j^2 + \lambda x m_j}  \label{nsterm}
\end{eqnarray}
\end{widetext}
The difficulty is the prefactor which introduces "interactions" between
the strings. Let us study it in two stages:

(i) {\it independent string approximation}. $Z(1,x)$ can be computed
exactly, integrating over momentum:
\begin{eqnarray} 
&&  \! \! \! Z(1,x) =  \int_{v>0} \frac{dv ~ v^{1/2}}{2 \pi \lambda^{3/2}}  dy Ai(y) 
\sum_{m =1}^\infty (-1)^m e^{\lambda m y - v m + \lambda x m} \nonumber 
\end{eqnarray}
where, as in \cite{dotsenko}, we used that for $\Re[w]>0$:
\begin{eqnarray}  \label{airytrick} 
&& \int_{-\infty}^\infty dy Ai(y) e^{y w} = e^{w^3/3}
\end{eqnarray} 
Rescaling $v \to \lambda v$, shifting $y \to y+v-x$ we obtain:
\begin{eqnarray} 
&&  Z(1,x) = -   \int_{v>0} \frac{dv ~ v^{1/2}}{2 \pi} dy Ai(y+v-x)  \frac{e^{\lambda y} }{1 + e^{\lambda y}} \nonumber
\end{eqnarray} 
after performing the sum. At large $\lambda$ the integration is only over $y>0$ and one obtains:
\be \label{limit}
 \lim_{\lambda \to \infty} Z(1,x) = - \int_{w>0} \frac{dw}{3 \pi}  w^{3/2} Ai(w - x)
\ee
Now we note that replacing in (\ref{nsterm}) $Z(n_s,x) \to Z(1,x)^{n_s}$ provides an {\it approximation} to the exact $g(x)$
{\it identical to setting the prefactor in (\ref{nsterm}) to unity}:
\begin{eqnarray}  
 && g_{ind}(x) = \exp(Z(1,x)) 
\end{eqnarray}
with (\ref{limit}) for $\lambda=\infty$ and $Prob_{ind}(f > x) = g_{ind}(x)$. One 
easily checks that this distribution $P_{ind}(f)$ is the one obtained in Ref. \cite{dotsenko}.
Indeed, the algebraic manipulations there are equivalent to setting the prefactor 
to unity at large $\lambda$ \cite{footnote4}. However this distribution  
has skewness $\gamma_1=0.96029$, incompatible with our numerics which shows instead for 
all $\lambda$ a {\it negative} skewness $\gamma_1^{F}$ bounded by (minus) the TW skewness
$\gamma_1=- 0.224084..$. Although one checks that it reproduces 
the leading tail for $f \to -\infty$ of the TW distribution, it differs from it. 
As we now show, including interactions between strings leads to TW.

(ii) {\it exact result for the generating function at any time}: 

We now derive an expression of $g(x)$ valid for any $\lambda$, in terms of a Fredholm determinant. Using the 
identity:
\begin{eqnarray}  \label{id}
&& det[ \frac{1}{i (k_i-k_j) \lambda^{-3/2} +  (m_i +m_j) } ] \\
&& = \prod_{i<j} \frac{(k_i-k_j)^2  + (m_i-m_j)^2 \lambda^3}{(k_i-k_j)^2 + (m_i+m_j)^2 \lambda^3} \prod_{i=1}^{n_s} \frac{1}{2 m_i} 
\end{eqnarray}
and manipulations as above, starting from (\ref{nsterm}) one finds:
\begin{eqnarray}
&& Z(n_s,x) = 
\int_{v_i>0}  \prod_{i=1}^{n_s} dv_i ~~ det[ K_x(v_i,v_j) ] \\
&& g(x) = Det[ 1 + P_0 K_x P_0 ] 
\end{eqnarray}
where $Det$ is a Fredholm determinant (FD) defined with integration on the real positive axis $\int_{v>0}$, i.e.
here and below we define $P_s$ the projector on $[s,+\infty[$. The kernel $K_x(v,v')=\Phi_x(v+v',v-v')$, where we
have defined the function:
\begin{equation}
\Phi_x(u,w) = - \int \frac{d k}{2 \pi} dy Ai(y + k^2 - x + u) \frac{e^{\lambda y - i k w}}{1+ e^{\lambda y}} 
\end{equation}
These formula generate the small $\lambda$ expansion but they are valid for all $\lambda$.

(iii) {\it free energy distribution in the large time limit} 

For large $\lambda$ one can replace
$\frac{e^{\lambda y }}{1+ e^{\lambda y}}  \to  \theta(y)$. Then one obtains for $\lambda=+\infty$:
\begin{eqnarray}
&& Prob(f > x) = g(x) = \det( 1 + P_{-\frac{x}{2}} \tilde K P_{-\frac{x}{2}} ) \\
&& \tilde K(v,v') = - \int_{y>0} \frac{d k}{2 \pi} dy Ai(y + k^2 +v+v') e^{- i k (v-v')} \nonumber
\end{eqnarray}
where all integrals in the FD are for $\int_{v > -x/2}$. One recovers in particular
$Z(1,x) = Tr K_x = \int_{v>-x/2} \tilde K(v,v)$ which yields (\ref{limit}) above. We can now use the following
identity between Airy functions \cite{identity}:
\begin{equation} \label{airyid} 
 \int dk Ai(k^2 + v + v') e^{i k (v-v')} = 2^{2/3} \pi Ai(2^{1/3} v) Ai(2^{1/3} v') 
\end{equation}
which immediately implies that:
\begin{eqnarray} 
\tilde K(v,v') = - 2^{1/3} K_{Ai}(2^{1/3} v, 2^{1/3} v')
 \end{eqnarray} 
where $K_{Ai}(v,v')= (Ai(v) Ai'(v') - Ai'(v) Ai(v'))/(v-v') = \int_{y>0} Ai(v+y) Ai(v'+y) $ is the Airy kernel.
Upon rescaling $v,v'$ by a factor $2^{-1/3}$ we obtain:
\begin{equation}
Prob(f > x=- 2^{2/3} s ) = Det(1 - P_{s} K_{Ai} P_{s}) = F_2(s)
\end{equation} 
i.e. the Tracy Widom distribution. Hence in the large time limit one recovers the
TW distribution. 

{\it (iv) free energy distribution for any time}

We now extract the free energy distribution for any time.
We use (\ref{extract}) which expresses the distribution of $\ln Z$ as
a convolution, i.e. $\ln Z = \lambda u_0 + \lambda u$ where $\lambda u_0$ is a unit Gumbel independent
random variable and the distribution of the variable $\lambda u=\ln \tilde Z$ is obtained as:
\begin{eqnarray}
&& p(u) =  \frac{1}{2 i \pi} ( Det(1+ P_0 K P_0) - Det(1+ P_0 K^* P_0) ) \nonumber \\
&& K(v,v') = \int \frac{d k}{2 \pi} dy 
Ai(y + k^2 + v+v') \frac{e^{\lambda y - i k (v-v')}}{e^{\lambda u} - i \epsilon - e^{\lambda y}} \nonumber \\
&& 
\end{eqnarray}
where all $\int_{v > 0}$ and $^*$ denotes complex conjugation. Using $1/(x-i \epsilon)=PV \frac{1}{x} + i \pi \delta(x)$, the complex kernel is written $K = K_1 + i K_2$ where, 
using (\ref{airyid}) one finds:
\begin{equation}
K_1(v,v') = 2^{\frac{1}{3}} PV  \int dy  \frac{Ai(2^{\frac{1}{3}} v
+ y )) Ai(2^{\frac{1}{3}} v'+ y ))}{e^{\lambda u- 2^{2/3} \lambda y} -1 }
\end{equation}
and
\begin{equation}
K_2(v,v') =  \frac{\pi}{2^{1/3} \lambda} P_{Ai}(2^{1/3} v + 2^{-2/3} u, 2^{1/3} v' + 2^{-2/3} u) 
\end{equation}
where $P_{Ai}(v,v')=Ai(v)Ai(v')$ is a rank one projector. The latter property
implies that $p(u)$ is a linear function of $K_2$ hence:
\begin{equation} \label{result}
p(u) = Det(1+ P_0 ( K_1 + \frac{\lambda}{\pi} K_2) P_0) - Det(1 + P_0 K_1 P_0) 
\end{equation}
which is our final expression \cite{footnotepos}. 

We can now compare our result with the very recent works \cite{spohnnewKPZ} on KPZ growth 
with the narrow wedge initial condition, to which our work also applies (see also \cite{math}). The correspondence reads that
$\frac{\lambda_{KPZ}}{2 \nu}h \equiv \ln Z$, $2 \nu \equiv T/\kappa$ and $D \lambda_{KPZ}^2 \equiv \bar c/\kappa^2$. There the distribution of $h$ was obtained, which translated here yields (up to an additive constant)
$\ln Z = \gamma \xi_t$ where $\gamma=2^{2/3} \lambda$ and the distribution of $\xi_t$ becomes identical to TW at large $t$. 
Upon rescaling of $v,v'$ by $2^{-1/3}$ our result (\ref{result}) can also be rewritten as:
\begin{eqnarray} \label{result}
&& 2^{2/3} p(u) = Det[1 - P_{2^{-\frac{2}{3}} u} (B_t - P_{Ai}) P_{2^{-\frac{2}{3}} u}] \nonumber \\
&& - Det[ 1 - P_{2^{-\frac{2}{3}} u} B_t P_{2^{-\frac{2}{3}} u} ] 
\end{eqnarray}
where $B_t$ is the kernel defined in \cite{spohnnewKPZ} hence the results coincide. 

Let us close by discussing the temperature dependence for experimentally relevant models, e.g.
either $R(u)$ with a finite range correlation, or a discrete model. For
any fixed $T$ one expects $\Delta F \equiv \overline{F^2}^{1/2}=A(T)
t^\theta$ at large $t$, with $\theta=\frac{1}{3}$. Concerning the {\it amplitude} $A(T)$ it is clear that
the $\delta$-function model reproduces only its high $T$ behaviour. 
Indeed, here we found $\Delta F=T f(\tilde t)$, with 
$f(\tilde t) \sim \tilde t^{1/4}$ at small $\tilde t$ from (\ref{logcum}) and 
$f(\tilde t) \sim \tilde t^{\theta}$ from our large $\lambda$ analysis. Hence large $\lambda$ yields the amplitude $A(T) \sim \kappa^{1/3} T^{-2/3}$ and this can only be interpreted as a high $T$ limiting behavior, i.e 
there is no way the $\delta$-function model can predict the amplitude for the distinct $T<T_{dep}$ regime, where it crosses over to a constant $A(0)$. This is illustrated in Fig. \ref{figf2} where $\sigma^2= \overline{(\ln z)^2}^c/(2^{4/3} \lambda^2)$ is plotted as a function of $T$, for increasing $t$. The fixed $T$ and large $t$ behavior is $\sigma^2 \sim A(T)^2 T^{4/3} \kappa^{-2/3}$, hence 
at low $T$ it behaves, for the discrete model as $\sim A(0)^2 T^{2/3}$ with a non-universal prefactor. At high $T$ we know from the small $\lambda$ prediction (\ref{logcum}) with $\bar c=1$ that it behaves as $\sim T^{2/3} t^{-1/6}$. For intermediate $T$ a plateau is thus predicted to develop. Its approach {\it from above} is described by the large $\lambda$ limit of the (universal) crossover function computed here. The low $T$ behaviour below the plateau is out of reach of the $\delta$-function model. The value of the plateau should equal the variance of the TW distribution $\sigma^2 = \sigma^2_{TW}= 0.81319..$. However one sees that the convergence is slow and requires very large polymer lengths \cite{us}. 

Similarly one surmises that $\Delta x \sim T^3 g(t/T^5)$
with $g(y) \sim y^{1/2}$ at small $y$ and $g(y)=y^\zeta$ at large $y$ interpolating between thermal diffusion $x \sim \sqrt{T t}$ and
$x \sim B(T) t^{2/3}$ with $B(T) \sim (\kappa T)^{-1/3}$ at high $T$, while for $T \lesssim T_{dep}$, 
$B(T) \approx B(T=0)$. A similar interpretation of Brunet's result for the winding $D \sim (\kappa T)^{-1/2}$ can be given. Note that
if the exponents were to assume their Flory values, $\zeta_F=\frac{3}{5}$, $\theta_F=\frac{1}{5}$, then high and low T regimes would merge without need for a plateau (i.e. $A(T)$ is constant at high $T$), and this is indeed the case in the mean field method \cite{mp}. 
This is not the case here, and this is because in the
 high $T$ regime more typical paths contribute, $\Delta F=O(T)$,
while the low $T$ problem is dominated by the lowest energy path, $\Delta F=O(1)$.

To conclude we have obtained the distribution of the free energy of directed polymers in the
high $T$ regime described by the attractive Lieb Liniger model, from the Bethe Ansatz. It becomes identical to the Tracy Widom distribution at large time, although the amplitudes exhibit
a distinct low temperature behavior. 

We thank E. Brunet, B. Derrida, V. Kazakov, H. Spohn, T. Sasamoto, and K. Zarembo for useful discussions. P. Calabrese thanks LPTENS for
hospitality. This work was supported by ANR grant 09-BLAN-0097-01/2.


\begin{thebibliography}{32}
\bibitem{footnoteconv}
Expression (\ref{nsterm}) can be defined as a series in $\lambda$ at fixed $X=\lambda x$. We use
everywhere the Airy representation (\ref{airytrick}) to define it at fixed $\lambda$.
\bibitem{footnotepos}
Note that $P(\tilde Z)$ and $p(u)$ are not necessarily positive functions, 
the only requirement is that after after convolution with the Gumbel distribution,
i.e. $\ln Z = \lambda u_0 + \lambda u$, the distribution of $\ln Z$ is positive
and normalized to unity.
\bibitem{footnote4}
the flaw in \cite{dotsenko} presumably originates from use
of undefined integrals, while here we use convergent ones.
\end{thebibliography}
\end{document}